# Language-integrated provenance in Haskell

## Jan Stolarek and James Cheney

a   LFCS, University of Edinburgh

**Abstract**    Scientific progress increasingly depends on data management, particularly to clean and curate data so that it can be systematically analyzed and reused. A wealth of techniques for managing and curating data (and its provenance) have been proposed, largely in the database community. In particular, a number of influential papers have proposed collecting provenance information explaining where a piece of data was copied from, or what other records were used to derive it. Most of these techniques, however, exist only as research prototypes and are not available in mainstream database systems. This means scientists must either implement such techniques themselves or (all too often) go without.

This is essentially a code reuse problem: provenance techniques currently cannot be implemented reusably, only as ad hoc, usually unmaintained extensions to standard databases. An alternative, relatively unexplored approach is to support such techniques at a higher abstraction level, using metaprogramming or reflection techniques. Can advanced programming techniques make it easier to transfer provenance research results into practice?

We build on a recent approach called *language-integrated provenance*, which extends language-integrated query techniques with source-to-source query translations that record provenance. In previous work, a proof of concept was developed in a research programming language called Links, which supports sophisticated Web and database programming. In this paper, we show how to adapt this approach to work in Haskell building on top of the Database-Supported Haskell (DSH) library.

Even though it seemed clear in principle that Haskell's rich programming features ought to be sufficient, implementing language-integrated provenance in Haskell required overcoming a number of technical challenges due to interactions between these capabilities. Our implementation serves as a proof of concept showing how this combination of metaprogramming features can, for the first time, make data provenance facilities available to programmers as a library in a widely-used, general-purpose language.

In our work we were successful in implementing forms of provenance known as where-provenance and lineage. We have tested our implementation using a simple database and query set and established that the resulting queries are executed correctly on the database. Our implementation is publicly available on GitHub.

Our work makes provenance tracking available to users of DSH at little cost. Although Haskell is not widely used for scientific database development, our work suggests which languages features are necessary to support provenance as library. We also highlight how combining Haskell's advanced type programming features can lead to unexpected complications, which may motivate further research into type system expressiveness.

## The Art, Science, and Engineering of Programming

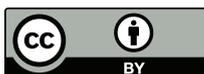

## 1 Introduction

Provenance is information (metadata) about the origin, derivation or history of an object. In computer systems, provenance is critical for assessing the reliability or trustworthiness of information, particularly in distributed settings where it is easy for plausible-looking information to be forged. As scientific progress increasingly depends on sharing data, provenance-tracking is an important requirement for scientific databases. For example, given a scientific data analysis pipeline, a scientist might spot some surprising results, and seek answers to provenance-questions such as:

- *why* was that record produced?
- *where* did the wrong-looking data values in it come from?
- *how* was the record constructed from the underlying data?

Due to the importance of relational databases for querying scientific data, there has been a great deal of research on provenance in databases (see surveys [6, 14]); some representative examples and systems include:

WHIPS [9], which recorded *lineage* information in the form of a set of input records explaining *why* that output record was produced;

DBNotes [2], which propagated annotations showing *where* some query results came from (were copied from) in the input;

ORCHESTRA [16], a data integration system that used a form of provenance explaining *how* some result records were obtained.

In this paper, we focus on where-provenance and lineage, as representative and very well-studied examples of provenance.

Previous work on provenance has led to the development of several systems, including those named above, implementing different provenance techniques. Most of these systems have been implemented as ad hoc extensions to mainstream database systems. More recently, systems such as Perm [13] showed that it is possible to support provenance by providing a middleware layer on top of the database, and translating queries to record their own provenance. However, there is still a significant 'impedance mismatch' between Perm's middleware layer and application needs.

The impedance mismatch problem between databases and programming languages has been addressed by *language integrated query*, an advance made in the last several years at the intersection of database and PL research [7, 12, 18]. As popularized by Microsoft's LINQ for .NET languages such as C# and F# [18], language-integrated querying treats queries as a typed, embedded domain-specific sublanguage (EDSL), making it much easier to safely connect to, query and update databases or other remote resources. Other approaches to language integrated queries include Links [8], a research Web programming language, and the Database-Supported Haskell system (DSH) [12, 22], a library that provides language-integrated queries for Haskell.

There is some previous work on defining provenance for general-purpose functional languages [1]. However, it emphasizes reasoning about security properties of provenance and does not consider provenance for language-integrated queries, which is our focus in this paper. Fehrenbach and Cheney [11] introduced a new approach (partly inspired by Perm), called *language-integrated provenance*. They showed that

**agencies**

| id | name | based_in | phone |
|---|---|---|---|
| 1 | EdinTours | Edinburgh | 412 1200 |
| 2 | Burns's | Glasgow | 607 3000 |

**externaltours**

| id | name | destination | type | price |
|---|---|---|---|---|
| 3 | EdinTours | Edinburgh | bus | 20 |
| 4 | EdinTours | Loch Ness | bus | 50 |
| 5 | EdinTours | Loch Ness | boat | 200 |
| 6 | EdinTours | Firth of Forth | boat | 50 |
| 7 | Burns's | Islay | boat | 100 |
| 8 | Burns's | Mallaig | train | 40 |

■ **Figure 1** Agencies and tours database

it is possible to implement the query transformations needed for where-provenance and lineage tracking at a high level in Links, before compiling queries to SQL. However, their approach required making nontrivial modifications to the Links interpreter. Moreover, Links is an experimental language, so while this work served as a proof of concept, it is still far from a solution to making provenance techniques practical.

We would like provenance to be more lightweight to implement, without requiring to modify a language or database implementation. Ideally, it would be a library that adds provenance tracking to database applications written in a mainstream language. As a step in this direction, we have extended DSH with support for provenance. Our design largely follows the approach taken in Links; the main contribution of this paper is showing how to overcome the engineering challenges of implementing provenance as (part of) a *library*, rather than a heavyweight *language extension*. We also highlight the strengths of Haskell for implementing provenance for a typed query EDSL, as well as challenges we encountered, particularly when transformations change not only the structure but also the *types* of the typed EDSL expressions. Our implementation is publicly available on GitHub: https://github.com/jstolarek/skye-dsh.

## 2 Background

Our contributions combine several streams of prior work on provenance, language-integrated query, and typed embedded domain-specific languages (EDSLs) in Haskell. In this section we present background needed for the rest of the paper.

### 2.1 Language-integrated query

Language-integrated query techniques build on the foundations of query languages for collections (lists, bags, sets) [4] and implemented in the Kleisli system [25]. These ideas influenced the Links system [8] and LINQ [18]; further discussion of the foundations of language-integrated query are given by Cheney, Lindley, and Wadler [7].

Language-integrated queries share common foundations with (list) *comprehensions* [4, 23] available in Haskell and an increasing number of other languages (Scala, C#, F#, Python). To illustrate, consider the sample database in Figure 1, which shows a table "agencies" containing information on travel agencies and a table

`"externaltours"` with information on tours organized by these agencies. A simple comprehension expression such as

```
[(e.name,a.phone) | e <- agencies
                  , a <- externaltours
                  , e.name == a.name
                  , e.type == "boat"]
```

would select the names and phone numbers of agencies offering boat tours. Using list semantics, this will return the result

```
[(EdinTours, 412 1200), (EdinTours, 412 1200), (Burns's, 607 3000)]
```

Such comprehensions correspond directly to conjunctive (select–from–where) SQL queries, and such queries can be extracted using a normalization algorithm [26] and then executed on the database. Extensions to comprehension syntax that handle more features from SQL, such as grouping and sorting, are also supported in Haskell [24].

## 2.2 Provenance

As explained in the introduction, there is a great deal of prior work on provenance in databases; we refer the interested reader to surveys [6, 14] for an overview. Tracking provenance allows the programmer to gain insight into how a query result is related to the underlying data. In this paper we will focus on tracking two forms of provenance:

**Where-provenance** provides information about the origin of values present in query result. Atomic values are identified by providing the name of the table together with the column name and row id where the value was copied from.

**Lineage** tells the programmer which database rows were used to produce a given row in the result.

Both forms of provenance can be implemented by *annotation propagation*: that is, conceptually speaking, by "tagging" each part of the input with an initial annotation (e.g. a unique identifier for that part), and then evaluating queries with a modified semantics that propagates the tags to the output in an appropriate way. For example, in the example query result earlier, we might ask for the where-provenance of the phone number for Burns's tour company, which would be an annotation (`agencies`, `phone`, 2) indicating that the phone number was copied from the `agencies` table, `phone` field, row 2. On the other hand, the lineage explaining why the row (`Burns's`, 607 3000) was produced consists of a *collection* of row references [(`agencies`, 2), (`externaltours`, 7)] indicating that both row 2 of the `agencies` table and row 7 of the `externaltours` table were needed to produce this output row.

Although we have been using relatively small example databases and queries, it is important to remember that the database tables might be large (e.g. thousands or millions of rows) and queries might be complex or opaque (e.g. generated by a program rather than written by a human). So, even though the "where" and "why" explanations are fairly easy to see for these small examples, automated support for computing these explanations is invaluable for more realistic settings.

Most research prototypes have provided this behavior by modifying the database implementation but it is also possible in principle to achieve the same effect by transforming queries to propagate their own provenance [2, 3, 13]. These ideas in turn inspired the language-integrated provenance approach of Fehrenbach and Cheney [11], which we follow in extending DSH with provenance tracking. This means that the query is transformed to track provenance information and only then compiled to SQL.

Where-provenance and lineage (and other forms of provenance) have most often been studied for relatively limited query languages, e.g. conjunctive or monotone SQL queries. While it is fairly straightforward to extend where-provenance to richer query languages, for lineage it is unclear what is the right semantics for some query operators, particularly so-called *non-monotone* operations such as set difference, grouping, or sorting. Since this is a general issue and orthogonal to the question of how to implement the query transformations, in this paper (as in [11]) we will restrict attention to a query language for which both forms of provenance are well-understood.

### 2.3 Haskell programming language

Haskell is a purely functional statically typed programming language. Its implementation in the Glasgow Haskell Compiler (GHC) offers a rich set of extensions that enhance expressiveness of Haskell's type system and enable meta-programming. These extensions make Haskell a good platform for implementing Embedded Domain Specific Languages (EDSLs).

We assume that the reader is familiar with some basic features of Standard Haskell: the standard Prelude, algebraic data types, type classes and list comprehensions. Readers unfamiliar with these features are encouraged to read Appendix A, which gives a brief overview of concepts essential to understanding the paper. We do not assume familiarity with non-standard language extensions offered by GHC. We present these below.

#### 2.3.1 Redefining standard Prelude and list comprehension syntax

GHC provides an extension `NoImplicitPrelude` that prevents implicit importing of the standard Prelude into a module. It allows the programmer to redefine basic functions and data types provided by the standard Prelude. This extension is complemented by `RebindableSyntax`, which allows user to override default desugaring of list comprehensions and some other built-in notations in Haskell. Together these extensions constitute a means to fully redefine the semantics of list comprehensions, which we shall demonstrate in Section 2.4.

#### 2.3.2 Generalized Algebraic Data Types (GADTs)

Generalized Algebraic Data Types (GADTs) are data types, whose type parameters (called *indices*) can be instantiated to concrete types that differ between constructors. Here is an example of a GADT that defines a simple language of expressions:

```
data Exp a where
    I   :: Int               -> Exp Int
    B   :: Bool              -> Exp Bool
    Add :: Exp Int -> Exp Int -> Exp Int
    Eq  :: Exp Int -> Exp Int -> Exp Bool
```

A value of type `Exp a` encodes an expression that evaluates to a value of type a. The `I` and `B` data constructors create integer and boolean expressions, respectively. `Add` allows to add two integers and `Eq` tests for integer equality. Notice how index variable a is instantiated to concrete type `Int` or `Bool` in type signature of each constructor. In this way one embeds simple typing rules for this expression language in the type system of Haskell. These rules make it impossible to pass a boolean value as an argument to addition or compare two booleans for equality. They also say that the result of addition is an `Int`, while the result of testing for equality is a `Bool` – just as specified by the indices in return types of `Add` and `Eq`. That approach to encoding typing rules of embedded language is a powerful tool offered by Haskell to authors of EDSLs.

### 2.3.3 Associated types and type families

In standard Haskell a type class defines a set of functions over a type. GHC's `TypeFamilies` extension allows to extend a type class with definitions of *associated type families* (also: *associated type synonyms*, *associated types*). A classical example is defining a type class of collections with an associated type family `Elem c` defining the type of the elements [5]:

```
class Collection c where
    type Elem c
    empty  :: c
    insert :: Elem c -> c -> c
```

When defining instances of `Collections` the programmer will need to say what `Elem` is:

```
instance Collection [e] where
    type Elem [e] = e
    ...
```

Type families can also be defined without being associated with a type class.[1] We typically think of type families as a form of type functions, i.e. functions taking types as arguments and returning types as a result. Type families become useful when one wants to express complex dependencies between types, especially when GADTs are involved.

### 2.3.4 Type proxies

Programming with type classes and type families carries a caveat. A polymorphic function in Haskell has type variables in its signature. At every call site of such a function, the compiler determines concrete instantiations for all type variables in a signature. But when a type variable appears only under type family applications or in type class constraints the compiler is unable to instantiate it.[2] To resolve this ambiguity, Haskell programmers resort to using additional arguments of a `Proxy` type:

---

[1] The idea of defining type families without associating them to a class has proven controversial. Readers interested in this aspect of Haskell's type system should take a look at [19].

[2] With the exception of *injective* type families [21]

```
data Proxy a = Proxy
```

Adding such an argument resolves a previously ambiguous type variable. It is now the caller's responsibility to provide a `Proxy` argument with a type signature of a desired type. `Proxy` arguments carry no runtime meaning since the type only takes a single value and exists only to guide the type checker.

### 2.3.5 Template Haskell

GHC comes with a powerful mechanism for meta-programming known as Template Haskell (TH) [20]. It allows programmers to write code that operates on Haskell's syntax trees. As such, TH can be used to implement repetitive fragments of an EDSL and provide a convenient interface for EDSL users. In our implementation we rely on TH to generate repetitive internal library code that handles tuples of lengths between 2 and 16. We also expose TH functions to the user so that all necessary boilerplate can be generated automatically. We stress that we use TH only for convenience – ours and users'. We could do without Template Haskell but that would make developing and using our implementation more tedious.

### 2.4 Database-Supported Haskell

Database-Supported Haskell (DSH) [12, 22] is an EDSL that allows Haskell programmers to write language-integrated queries. It relies on Haskell extensions described in Section 2.3 to override default desugaring of list comprehensions and compile these comprehensions to SQL queries that are executed on a database. In this section we provide a quick tour of DSH as seen from the user's perspective. In Section 3 we will go into implementation details.

To illustrate, we continue with the running example data (Figure 1) and query discussed earlier. In order to write DSH queries on this database one first needs to declare data types that correspond to table schemas. For the `agencies` table this data type will look like this:

```
data Agency = Agency { a_id       :: Integer
                     , a_name     :: Text
                     , a_based_in :: Text
                     , a_phone    :: Text }
```

Each of the record fields corresponds to a column in a database table. Having the `Agency` data type defined allows us to write a database table declaration[3]:

```
agencies :: Q [Agency]
agencies = table "agencies"
              ["a_id", "a_name", "a_based_in", "a_phone"]
              (TableHints [Key ["a_id"]])
```

---

[3] To simplify our presentation we prettify some of our code snippets. The actual code in the implementation is slightly different but this is only to elide irrelevant technical details.

A table declaration consists of the name of the database table (`"agencies"`), a non-empty list of table columns and table hints that contain a list of table keys (at least one key must be declared, compound keys are permitted). The type of `agencies` is declared to be a list of `Agency` values stored inside `Q`, which is a type used by DSH to wrap all values coming from queries. We also need record selectors that work on values in `Q`, e.g.:

```
a_nameQ :: Q Agency -> Q Text
```

These selectors and other boilerplate are generated by DSH using Template Haskell. All the user needs to write is:

```
deriveDSH              ''Agency
generateTableSelectors ''Agency
```

We also need an analogous declaration for the `externaltours` table: an `ExternalTours` data type to represent table row, accessors prefixed with `et_` rather than `a_`, `externaltours` table declaration, and TH calls to generate boilerplate.

With these declarations available, the programmer can use list comprehension notation to write a query that returns names and phone numbers of all agencies organizing boat tours:

```
q1 :: Q [(Text, Text)]
q1 = [ tup2 (et_nameQ et) (a_phoneQ a)
     | a  <- agencies
     , et <- externaltours
     , a_nameQ a == et_nameQ et
     , et_typeQ et == "boat" ]
```

Interesting things to note in this query are:

- `agencies` and `externaltours` table declarations can be iterated over as if they were lists. Indeed, this is what we declared their types to be — lists of either `agencies` or `externaltours` values.
- all record fields are projected using `Q`-aware accessors
- a tuple value is constructed using the `tup2` smart constructor, again to ensure proper handling of types inside `Q`.

DSH compiles the above list comprehension to the following SQL query:

```
SELECT a1.et_name AS i1, a0.a_phone AS i2
FROM agencies AS a0, externaltours AS a1
WHERE (a0.a_name = a1.et_name) AND (a1.et_type = 'boat')
```

## 3    Design of provenance tracking in DSH

Following the approach in [3], we treat provenance information as additional metadata attached to results of a query. A crucial requirement for provenance is that it cannot be forged: the programmer should be allowed neither to create provenance information out of thin air nor to take existing provenance and re-attach it to data it does not

| id | name | phone (with where-provenance) | |
|---|---|---|---|
| 1 | EdinTours | (data = 412 1200, | prov = ("agencies", "a_phone", 1)) |
| 2 | EdinTours | (data = 412 1200, | prov = ("agencies", "a_phone", 1)) |
| 3 | Burns's | (data = 607 3000, | prov = ("agencies", "a_phone", 2)) |

■ **Figure 2**  Example query results with where-provenance tracking belong to. The implementation of provenance in Links, on which we model our development, partially enforces these security requirements, but once provenance leaves the query it becomes ordinary data that programmer can manipulate [11]. Our design goal is to implement the requirements defined in [3] fully. Another design goal is to provide programmers with the ability to easily extend existing queries with where-provenance and lineage tracking. To this end we have designed an interface that seamlessly integrates with the DSH library.

In this section we present an overview of what the provenance interface looks like and explain how it meets the above requirements. We also define where-provenance and lineage transformations in a theoretical setting. Both of these transformations work by rewriting queries.

### 3.1 Provenance tracking in DSH queries by example

Continuing the example from Section 2.4 assume that the programmer thinks some obtained phone numbers are incorrect. She might elect to track where-provenance for the `"phone"` field of `"agencies"` table. In our system this requires modifying the original definition of `Agency` data type by annotating the relevant field with a where-provenance annotation:

```
data AgencyWP = AgencyWP { a_id      :: Integer
                         , a_name    :: Text
                         , a_based_in :: Text
                         , a_phone   :: WhereProv Text Integer }
```

where `WhereProv` is a data type of where-provenance annotations, `Text` is the field type and `Integer` is the type of primary key used to identify table rows.

With this modified definition value of `a_phone` field now consists of a data component, that stores the actual field value, and a provenance component, that stores information about table, column, and row key that uniquely identify the origin of a value in the data base. These components can be projected using `dataQ` and `provQ` helper functions, respectively. Since we have changed the type of `a_phone` field, the type of our example query q1 changes from `Q [(Text, Text)]` to `Q [(Text, WhereProv Text Integer)]`. The query itself remains unchanged. The result of running our query with where-provenance tracking is shown in Figure 2. If we wanted to write a query that behaves just as our original q1 query, i.e. returns a phone number without a provenance annotation, we would need to explicitly project the data component when constructing the result tuple:

```
q1' :: Q [(Text, Text)]
```

| id | name | phone | lineage |
|---|---|---|---|
| 1 | EdinTours | 412 1200 | [ ( "agencies", 1 ),( "externaltours", 5 )] |
| 2 | EdinTours | 412 1200 | [ ( "agencies", 1 ),( "externaltours", 6 )] |
| 3 | Burns's | 607 3000 | [ ( "agencies", 2 ),( "externaltours", 7 )] |

■ **Figure 3**  Example query results with lineage tracking

```
q1' = [ tup2 (et_nameQ et) (dataQ (a_phoneQ a))
      | {- unchanged from q1 -} ]
```

Looking at the results of the original `q1` query, the programmer might have also realized that one of the resulting rows is different than expected, for example because she thinks a given agency does not organize boat trips. In such a situation tracking lineage allows to obtain information which database rows were used to construct each row of the result. Since lineage information is attached to result rows rather than to individual fields, there is no need to modify `Agency` data type definition. To track the lineage of a query, the user needs to call a `lineage` library function on the original query and modify the return type accordingly:

```
q1'' :: Q (LT [(Text, Text)] Integer)
q1'' = lineage (Proxy :: Proxy Integer) q1
```

Here `LT` indicates that a lineage annotation has been attached to a query, `[(Text, Text)]` is the type of query and `Integer` is a type of row keys. The `lineage` function also needs a proxy argument to disambiguate the type of keys. The result of running our example query with lineage tracking is shown in Figure 3. Each row of the `q1''` result now consists of a data component and a lineage component, which can be projected using `lineageDataQ` and `lineageProvQ` library functions. Every lineage annotation consists of a list of table name and row key pairs. Since we are identifying whole rows, there is no need to track information about columns. Our implementation of lineage also requires that table definitions explicitly contain functions for projecting primary key values from a row. Details are given in Section 4.4.

Appendix B gives full code for our examples.

## 3.2  Surface encoding of where-provenance

Tracking where-provenance requires the ability to uniquely describe locations in a database. This is achieved with the `WhereProvAnnot` data type that stores table name, column name and row key:

```
data WhereProvAnnot k = WhereProvAnnot
    { where_prov_table  :: Text
    , where_prov_column :: Text
    , where_prov_key    :: k }
```

Notice that `WhereProvAnnot` is polymorphic in the row key type `k`. Unlike the original approach in Links [11], in our implementation rows can be identified using any type of key, including a compound one, and not just integers.

Per [11], values annotated with where-provenance can have a concrete provenance value indicating that it originates from the database. We also allow values to have blank provenance annotations, denoted with the '⊥' symbol, indicating the value came from the query itself.[4] We achieve this by defining data type `WhereProv`:

```
data WhereProv a k where
    NoProv    :: BasicType a => a -> WhereProv a k
    WhereProv :: BasicType a => a -> WhereProvAnnot k -> WhereProv a k
```

The `WhereProv` data constructor attaches a where-provenance annotation to a value; `NoProv` represents blank provenance. The `BasicType` constraint ensures that where-provenance can only be attached to primitive values that can be stored in a single database cell.

To allow programmers to access data and provenance components of where-provenance annotated values, we create two helper functions `dataQ` and `provQ`, respectively. We also provide `emptyProvQ`, which attaches blank where-provenance to a value. Importantly, the exposed library interface ensures that `WhereProv` and `WhereProvAnnot` data types remain abstract and programmers cannot create their values by hand. Moreover, it is not possible to take an existing where-provenance and attach it to another value. These features combined together fulfill the crucial requirement that provenance can only be inspected but not forged.

### 3.3 Surface encoding of lineage

Tracing lineage requires attaching a set of lineage entries to each row of a query result. A single lineage entry consists of table name and row key, thus uniquely identifying a row in a database. We define:

```
data LineageAnnotEntry k = LineageAnnotEntry
    { lineage_table :: Text, lineage_key    :: k }

data Lineage a k where
    Lineage :: a -> Set (LineageAnnotEntry k) -> Lineage a k
```

Data and provenance components of lineage-annotated rows can be accessed using `lineageDataQ` and `lineageProvQ` functions. Empty lineage can be attached to a row using `emptyLineageQ`. Just like with where-provenance, the programmer can inspect lineage but is prevented from creating it by hand or reattaching existing lineage to values it does not belong to, thus ensuring that safety requirements for provenance are fulfilled.

### 3.4 Core calculus for provenance in DSH

Figure 4 presents types and syntax of a core calculus we have developed to model DSH's Frontend Language (FL), a language into which the library desugars list com-

---

[4] Provenance calculus in [11] guarantees that all provenance annotations are non-blank using the type system, but most work on where-provenance in databases [2, 3] allows blank annotations.

| Primitive types | $\tau$ | $::=$ | **()** \| **Bool** \| **Int** \| **String** |
|---|---|---|---|
| QA types | $\delta$ | $::=$ | $\tau \mid [\delta] \mid (\delta_1, \ldots, \delta_n) \mid \tau^W \mid [\delta^L]$ |
| Key types | $\theta$ | $::=$ | $\tau \mid (\tau_1, \ldots, \tau_n)$ |
| Types | $\sigma$ | $::=$ | $\delta \mid \sigma_1 \rightarrow \sigma_2$ |
| Row types | $R$ | $::=$ | $\cdot \mid R, l : \tau$ |

| Where-provenance annot. | $W$ | $::=$ | $\bot \mid (t : \textbf{String}, c : \textbf{String}, k : \theta)$ |
|---|---|---|---|
| Lineage annotations | $L$ | $::=$ | $\bot \mid (t : \textbf{String}, k : \theta) \mid L_1 \oplus L_2$ |
| Provenance annotations | $A$ | $::=$ | $W \mid L$ |
| Built-in functions | $F$ | $::=$ | **concatMap** \| **map** \| **append** \| **reverse** |
| | | \| | **guard** \| **cons** \| **zip** \| $M.n$ |
| Expressions | $M$ | $::=$ | $c \mid x \mid () \mid [M_1, \ldots, M_n] \mid (M_1, \ldots, M_n)$ |
| | | \| | $\lambda x.M \mid M\ M_1 \ldots M_n \mid \textbf{table}_{(t : \textbf{String}, \phi : R \rightarrow \theta)}$ |
| | | \| | $M^A \mid M.\textbf{data} \mid M.\textbf{prov}$ |

■ **Figure 4** Frontend Language types and syntax

prehensions (Section 4.2 provides details about FL). Our calculus has the following important features and differences from FL originally implemented in DSH:

- Primitive types include unit, booleans, integers and strings. These correspond to types that can be stored in database cells. FL supports several other primitive types that correspond to data types supported by database engines like floats and dates. We omit those to simplify our presentation.

- QA types are types of expressions that can be desugared from surface Haskell into FL. Primitive types can be annotated with where-provenance. Elements of collections can be annotated with lineage information.

- Keys can be either primitive or compound, the latter being represented as tuples.

- Row types correspond to user-defined Haskell records like `Agency` in our example in Section 2.4. Records are absent in actual FL – they are translated to tuples. We use them however in our calculus to represent tables.

- $\bot$ represents empty provenance, i.e. provenance of values that were created by a query rather than being extracted from a database.

- Lineage annotations consist of multiple entries that are appended using $\oplus$.

- DSH supports nearly sixty built-in primitive functions that correspond to various primitives provided by database engines. In our calculus we significantly limit the number of supported built-ins. The primary reason for this is that for omitted primitives it is generally not known how to track provenance, usually due to non-monotonicity. Handling these primitives remains an open research problem that is beyond the scope of this paper.

- $M.n$ built-in function represents tuple projections, where $n$ is number of component to project.

- Tables are identified by name $t$ and contain a function $\phi$ that allows to project the primary key from a row.

$$\begin{aligned}
\mathfrak{W}(\mathbf{table}_{(n:\mathbf{String}, \phi:R\to\theta)}) &= \mathbf{map}\ (\lambda x. \mathfrak{W}_{t,\phi,x}(x))\ \mathbf{table}_{(t:\mathbf{String}, \phi:R\to\theta)} \\
\mathfrak{W}_{t,\phi,x}(\cdot) &= \cdot \\
\mathfrak{W}_{t,\phi,x}(R, l:\tau) &= \mathfrak{W}_{t,\phi,x}(R), l \\
\mathfrak{W}_{t,\phi,x}(R, l:\tau^W) &= \mathfrak{W}_{t,\phi,x}(R), l^{(t,"l",\phi(x))}
\end{aligned}$$

■ **Figure 5** Where-provenance transformation for tables

- Expressions can be annotated with provenance annotations (either where-provenance or lineage), which is represented by superscript $A$ in the expression $M^A$. From every annotated expression we can project out its data and provenance component. Following the original approach in [11] our system does not allow to compose where-provenance and lineage tracking in the same query.

It is also important to distinguish this core calculus from the "core calculus for provenance" proposed by Acar, Ahmed, Cheney, and Perera [1]. In that paper, a pure functional language was extended with different provenance-tracking semantics. However, this language did not include language-integrated query capabilities and the emphasis was on reasoning about security properties of provenance, not implementing provenance tracking for queries as is our goal here.

### 3.5 Where-provenance transformation

Where-provenance transformation $\mathfrak{W}$ is defined in Figure 5. It rewrites table expressions by calling the $\mathfrak{W}_{t,\phi,x}$ helper transformation on each row, which extends data pulled from a database with where-provenance annotations as necessary. In the third equation of $\mathfrak{W}_{t,\phi,x}$ a field marked with where-provenance tracking receives an annotation that contains the table name $t$, column's name "$l$", and row's key $\phi(x)$. It is sufficient to rewrite table expressions only, because where-provenance is explicit in types (cf. examples in Section 3.1) and the programmer is forced to write well-typed queries that take account of provenance annotations so that the code passes Haskell's type checking.

### 3.6 Lineage transformation

The lineage transformation is defined by translation of terms (Figure 6) and a corresponding translation of types (Figure 7). Transformation $\mathfrak{L}_\theta$ is parametrized by the type of table keys $\theta$. This signifies that we support polymorphic keys but require that all tables used in a query have the same type of primary key.

The two most important equations of the transformation are for **table** and **concatMap** expressions. The equation for **table** says that whenever we pull data from a database we annotate each row with information about its origin (table name and key). These annotations act as meta-data that is propagated through a query. A crucial part of that propagation is performed by **concatMap**, because list comprehensions desugar to **concatMap** calls. The principal idea behind the transformation for **concatMap** is this. Each element of the xs list can have lineage associated with it, obtained via a recursive call $\mathfrak{L}_\theta(xs)$. Then for every element of this list we call function $\lambda f.M$,

$$\mathfrak{L}_\theta(\mathbf{table}_{(n:\mathbf{String},\phi:R\to\theta)}) = \mathbf{map}\ (\lambda x.x^{(t,\phi(x))})\ \mathbf{table}_{(t:\mathbf{String},\phi:R\to\theta)}$$

$$\mathfrak{L}_\theta(\mathbf{concatMap}\ (\lambda f.M)\ xs) = \mathbf{concatMap}\ (\lambda x.$$
$$\mathbf{map}\ (\lambda z.(z.\mathbf{data}^{z.\mathbf{prov}\oplus x.\mathbf{prov}}))\ \mathfrak{L}_\theta((\lambda f.M)(x.\mathbf{data})))$$
$$\mathfrak{L}_\theta(xs)$$

$$\mathfrak{L}_\theta(\mathbf{map}\ (\lambda f.M)\ xs) = \mathbf{concatMap}\ (\lambda x.$$
$$\mathbf{map}\ (\lambda z.(z.\mathbf{data}^{z.\mathbf{prov}\oplus x.\mathbf{prov}}))\ \mathfrak{L}_\theta[(\lambda f.M)(x.\mathbf{data})])$$
$$\mathfrak{L}_\theta(xs)$$

$$\mathfrak{L}_\theta(\mathbf{append}\ xs\ ys) = \mathbf{append}\ (\mathfrak{L}_\theta(xs))\ (\mathfrak{L}_\theta(ys))$$

$$\mathfrak{L}_\theta(\mathbf{reverse}\ xs) = \mathbf{reverse}\ (\mathfrak{L}_\theta(xs))$$

$$\mathfrak{L}_\theta(\mathbf{zip}\ xs\ ys) = \mathbf{map}\ (\lambda x.(x.1.\mathbf{data},x.2.\mathbf{data})^{x.1.\mathbf{prov}\oplus x.2.\mathbf{prov}})$$
$$(\mathbf{zip}\ (\mathfrak{L}_\theta(xs))\ (\mathfrak{L}_\theta(ys)))$$

$$\mathfrak{L}_\theta(c) = c$$

$$\mathfrak{L}_\theta(x) = x$$

$$\mathfrak{L}_\theta([M_1,\ldots,M_n]) = \mathbf{map}\ (\lambda x.x^\perp)\ [\mathfrak{L}_\theta(M_1),\ldots,\mathfrak{L}_\theta(M_n)]$$

$$\mathfrak{L}_\theta(\mathbf{cons}\ x\ xs) = \mathbf{cons}\ (\mathfrak{L}_\theta(x)^\perp)\ (\mathfrak{L}_\theta(xs))$$

$$\mathfrak{L}_\theta(\mathbf{guard}\ b) = \mathbf{map}\ (\lambda x.x^\perp)\ (\mathbf{guard}\ b)$$

$$\mathfrak{L}_\theta(M_1,\ldots,M_n) = (\mathfrak{L}_\theta(M_1),\ldots,\mathfrak{L}_\theta(M_n))$$

$$\mathfrak{L}_\theta(M.n) = (\mathfrak{L}_\theta(M)).n$$

■ **Figure 6** Lineage transformation

$$\mathfrak{L}_\theta(\mathbf{()}) = \mathbf{()} \qquad \mathfrak{L}_\theta(\mathbf{String}) = \mathbf{String}$$
$$\mathfrak{L}_\theta(\mathbf{Bool}) = \mathbf{Bool} \qquad \mathfrak{L}_\theta([\delta]) = [\mathfrak{L}_\theta(\delta)^L]$$
$$\mathfrak{L}_\theta(\mathbf{Int}) = \mathbf{Int} \qquad \mathfrak{L}_\theta(\delta_1,\ldots,\delta_n) = (\mathfrak{L}_\theta(\delta_1),\ldots,\mathfrak{L}_\theta(\delta_n))$$

■ **Figure 7** Lineage type translation

remembering that we need to pass as an argument only the data component. The result of $((\lambda f.M)(x.\mathbf{data}))$ is a list, each of whose elements can have a lineage annotation assigned to it. Thus another recursive call to $\mathfrak{L}_\theta$. In this way every element of the resulting list has two lineages associated with it: one from $\mathfrak{L}_\theta(xs)$ and another from $\lambda f.M$. These lineages need to be appended together, which is performed in the innermost lambda.

The equation for **map** is very similar, except that we wrap the result of $((\lambda f.M)$ $(x.\mathbf{data}))$ into a singleton list before passing it to $\mathfrak{L}_\theta$. The treatment of **append**, **reverse**, tuples and projections is straightforward. The **zip** case is a bit more involved, as it requires restructuring data and provenance components accordingly. Since lineage is attached only to data pulled from the database, we don't assign any lineage to constants and variables (but variables have their type transformed according to Figure 7). For list literals (**list** and **cons** cases) we assign empty lineage to every element to match the type translation defined in Figure 7, but also call $\mathfrak{L}_\theta$ recursively.

The **guard** operation supports a form of filtering: if its Boolean argument is true, it returns a singleton collection `[()]`, otherwise the empty sequence `[]`. Comprehensions involving filtering are translated to FL code using the **guard** construct. Since boolean expressions have no associated lineage (per Figure 7), the lineage of the result (in the **true** case) is empty. One might expect that we should track lineage of condition $b$ and assign it to the whole **guard** $b$ expression, but this is not considered in lineage (at least for monotone queries).

Appendix C contains examples that demonstrate how rules from Figure 6 work in practice to extend a query with lineage tracking.

## 4 Implementation

In this section we explain our implementation of provenance transformations, how they fit into the DSH compilation pipeline, and the technical challenges encountered and overcome in the process.

### 4.1 Database Supported Haskell architecture

The DSH query compilation pipeline consists of several stages:

1. **Haskell source.** To enable the DSH library, the programmer needs to import DSH modules and enable GHC's language extensions that allow overriding of GHC's default list comprehension desugaring. Now queries get desugared into *Frontend Language*.

2. **Frontend Language (FL).** FL syntax tree is defined as a GADT, where type indices enforce typing invariants. FL is translated into Comprehension Language.

3. **Comprehension Language (CL).** CL expressions are explicitly typed, i.e. each expression stores its own type. Types are recovered from the structure of FL expressions. CL is used primarily for optimizations.

4. **Additional translations.** DSH employs several additional intermediate representations that translates a query expression to one or more flat relational queries. Since our implementation does not affect or depend on the details of these stages, we consider them as a black box.

5. **Relational algebra and SQL.** The final step generating SQL queries is performed by a backend library that is external to DSH. This allows DSH to work with different SQL backends.

We perform provenance transformations when a query is still represented in Frontend Language (for lineage) or during translation from FL to Comprehension Language (for where-provenance). For this reason we focus our attention on FL.

### 4.2 DSH user interface and the Frontend Language

The DSH library interface exposes several type classes, Template Haskell helper functions and a special type `Q`. These allow users to write list comprehensions that desugar into FL. At the heart of FL is the `Exp` data type, that represents the abstract syntax tree of a desugared list comprehension. Below is a fragment of the definition of `Exp`:

```
data Exp a where
    BoolE      :: Bool -> Exp Bool
    ListE      :: Reify a => [Exp a] -> Exp [a]
    TableE     :: Reify a => Table -> Exp [a]
    AppE       :: Fun a b -> Exp a -> Exp b
    VarE       :: Reify a => Integer -> Exp a
    TupleConstE :: TupleConst a -> Exp a
```

`Exp` is a GADT with constructor indices encoding typing invariants. Constants are indexed directly by their type (e.g. `BoolE`). List expression `ListE` stores a list of expressions of type a and is indexed by `[a]`. Table expression `TableE` stores table meta-data and, like `ListE`, is indexed by `[a]`. The definition of application relies on data type `Fun`, that lists all built-in primitive functions. Each function is indexed by its input and output type. Applications constructed with `AppE` ensure that the type of the argument matches that required by a function. Variables `VarE` are internally represented with a unique integer identifier and can have any type, hence their index is not constrained. Tuple expressions rely on auxiliary data type `TupleConst`:

```
data TupleConst a where
    Tuple2E :: (Reify t1, Reify t2)
            => Exp t1 -> Exp t2 -> TupleConst (t1, t2)
    -- data constructors up to Tuple16E
```

An important observation here is that `TupleConst` is also a GADT with indices encoding a tuple's arity and its component types, but this information is lost when `TupleConst` is wrapped in `TupleConstE` in `Exp`.

FL also has a GADTs for internal representation of types:

```
data Type a where
    BoolT  :: Type Bool
    ListT  :: Type a -> Type [a]
    ArrowT :: Type a -> Type b -> Type (a -> b)
    TupleT :: TupleType a -> Type a

data TupleType a where
    Tuple2T :: Type t1 -> Type t2 -> TupleType (t1, t2)
    -- data constructors up to Tuple16T
```

The `Type` will become important in the implementation of lineage transformation.

Some data constructors in `Exp` and all in `TupleConst` restrict their indices to belong to a `Reify` type class that provides a single method `reify`:

```
class Reify a where
    reify :: a -> Type a
```

Translation of FL into CL requires that it is possible to reconstruct the type of every FL expression. For the majority of expressions it is possible to reconstruct their type just from the syntax tree structure. For some expressions though this is impossible and extra guidance is required. That's when the `Reify` type class constraint is used.

Central to conversion between surface Haskell and Frontend Language is the `QA` type class:

```
class (Reify (Rep a)) => QA a where
    type Rep a
    toExp :: a -> Exp (Rep a)
    frExp :: Exp (Rep a) -> a
```

Types that are instances of `QA` can be converted from surface Haskell to an internal representation in FL. The associated type family `Rep` defines how a surface Haskell type is represented in FL. For primitive types this is an identity function, i.e. `Bool` is represented as `Bool`, and so on. Data types defined to represent table rows, like `Agency` in our example in Section 2.4, are represented as tuples of arity equal to the number of the data type's fields. For example `Agency` data type is represented internally as a four-tuple (`Integer`, `Text`, `Text`, `Text`). The functions `toExp` and `fromExp` are used to convert between surface and internal representations of a data type. Importantly, the internal expression `Exp` representing a value of type a is indexed with `Rep a`. The (`Reify (Rep a)`) constraint on `QA` enforces that internal representation types can be typed in the Frontend Language. Instances of `QA` are generated automatically by `deriveDSH` Template Haskell function.

In our examples in Section 2.4 we saw that all DSH expressions are wrapped in the `Q` type:

```
newtype Q a = Q (Exp (Rep a))
```

Expressions constructed when writing DSH queries are values of `Exp`, but wrapping them in `Q` allows the programmer to conveniently operate on surface types and not their internal representations.

### 4.3 Implementing where-provenance

Implementing the where-provenance transformation described in Section 3.5 requires us to transform `TableE` expressions whenever a where-provenance annotation is present on at least one of the table's columns. This raises the first implementation question: how to detect the presence of where-provenance annotations? As shown in Section 3.2 data type declarations that map to database rows have their fields tagged with the `WhereProv` data type. Unfortunately there is no way to access this information inside the DSH compilation pipeline. Thus we have adopted a different solution, where we extend DSH table declarations with where-provenance information:

```
agenciesWP :: Q [AgencyWP]
agenciesWP = table "agencies"
    ["a_id", "a_name", "a_based_in", "a_phone"]
    (TableHints [Key ["a_id"]] (WhereProvenance ["a_phone"]))
```

Here `WhereProvenance` is an extra table hint that lists all columns that should have their where-provenance tracked. If there is no provenance tracking for a given table, the programmer has to supply a `NoProvenance` hint. During translation from FL to CL `TableE` expressions that contain provenance hints are extended with where-provenance information. Taking query `q1'` as an example, the where-provenance transformation is logically equivalent to rewriting `a <- agencies` generator in the following way:

```
q1W :: Q [(Text, Text)]
q1W = [ tup2 (et_nameQ et) (dataQ (a_phoneQ a))
      | a <- [ agencyWP (a_idQ a') (a_nameQ a') (a_based_inQ a')
                 (WhereProv a' (WhereProvAnnot "agencies" "a_phone"
                                                (a_id a')))
             | a' <- agencies ]
      , {- unchanged from q1 -} ]
```

where `agencyWP` is a smart constructor for `AgencyWP` data type. Note however that the programmer would not be able to write such a transformation by hand because `WhereProv` and `WhereProvAnnot` constructors remain abstract, as explained in Section 3.2.

A crucial element that makes such an implementation possible is that a `table` declaration can be assigned to be a list of elements of *any* type. This is possible even if the structure of the assigned type does not match the structure of the actual database table. This is precisely what we are doing when handling where-provenance, since the structure of data types with provenance annotations does not correspond to the structure of the database. In our example above we declare `agenciesWP` to be a list of `AgencyWP`, even though the database table does not store where-provenance data for `a_phone` field. Additional rewriting steps performed when translating `TableE` expressions to CL ensure that this mismatch is eliminated. At the same time when we define a table to contain a where-provenance annotated data type, Haskell's type system will ensure that the query over that table takes additional provenance meta-data into account.

A huge engineering win of the where-provenance transformation is that it integrates seamlessly with existing compilation pipeline. The implementation is concise and limited to a single place in the DSH source code.

### 4.4 Implementing lineage

Based on experience with implementing where-provenance, we were hoping to implement lineage transformation in a similar fashion, i.e. using extra annotations to trigger query rewriting during compilation from FL to CL. That however turned out not to be possible. Recall that a query written in list comprehension form is desugared into an FL syntax tree. That tree has to be a well-typed Haskell expression, before being passed further down the compilation pipeline. This was not an issue for the where-provenance annotated query. We assigned a provenance-annotated type to a `table` declaration and it was enough to make the syntax tree well-typed at the surface level. Later on in the compilation pipeline we amended the structure of `TableE` expressions to actually match the type it was assigned.

We cannot use this trick with lineage since it is not a local transformation of a single expression, but a complex rewriting of the whole syntax tree. Thus we have to implement lineage transformation as an FL-to-FL pass that takes a well-typed syntax tree representing a query without lineage and returns a transformed syntax tree representing a query with lineage tracking. This transformation is exposed to the programmer as a library function, as demonstrated in Section 3.1.

Implementing the lineage translation according to rules given in Figures 6 and 7 proved to be challenging. The main problem we face is implementing the type translation in Figure 7 and ensuring that it is consistent with the expression translation in Figure 6. In DSH, types are encoded at several different levels in several different ways. Firstly, there are *surface types* present in Haskell source code when writing queries. These include user-defined data types, like `Agency`, but also library-exposed types, like `Lineage` and `LineageAnnotEntry`, defined in Section 3.3. Secondly, there are *internal types*, which are Haskell types used in FL to index the `Exp` GADT. Surface types are translated to internal types by the `Rep` type family. Finally, there are explicit type representations embedded in CL syntax trees and represented as Haskell expressions using DSH's `Type` GADT. For each of those representations we need to separately implement type translation from Figure 7 and then convince GHC's type checker that all three implementations are equivalent. Sections 4.4.1 and 4.4.2 provide the details. Readers wishing to omit technical discussion of Haskell's type-level programming may skip directly to Section 4.4.3.

### 4.4.1 Implementing lineage type translation

We implement translation of surface types using an `LT` type family, already demonstrated in an example in Section 3.1:

```
type family LT a k
type instance LT Bool   k = Bool
type instance LT [a]    k = [Lineage (LT a k) k]
type instance LT (a, b) k = (LT a k, LT b k)
```

where a is a type to which we are attaching lineage and `k` is the type of key used for identifying table rows. Equations of `LT` correspond to equations in Figure 7, except that we only show here a single equation for handling pairs and primitive types. The most interesting equation is for lists. It calls `LT` recursively on the type of a and annotates the result with lineage. We make `LT` an *open type family* because the set of surface types is open. The programmer is expected to define new types, like `Agency`, and each new type will add a new equation to the definition of `LT`:

```
type instance LT Agency k = Agency
```

Implementing type translation for internal types requires another type family:

```
type family LineageTransform a k where
    LineageTransform Bool   k = Bool
    LineageTransform [a]    k = [(LineageTransform a k, [(Text, k)])]
    LineageTransform (a,b) k =
        (LineageTransform a k, LineageTransform b k)
```

As previously, a is a type to which we are attaching lineage, k is the type of keys, and we only show one equation for primitive types and one for tuples. We make `LineageTransfrom` a *closed type family* [10] because the set of internal types is predefined in the library and cannot be extended by the user.

As previously, the most interesting equation in `LineageTransform` is the one for lists, especially the `[(Text, k)]` bit in the right-hand side. It originates from the internal representation of a lineage annotation defined using the `Rep` type family:

```
instance (QA k) => QA (LineageAnnot k) where
    type Rep (LineageAnnot k) = [(Text, Rep k)]

instance (QA a, QA k) => QA (Lineage a k) where
    type Rep (Lineage a k) = (Rep a, Rep (LineageAnnot k))
```

The surface type `Lineage a k` is thus internally represented by a syntax tree of type `Exp (a, [(Text, k)])`.

For our lineage transformation code to work we have to show that we can convert between lineage-annotated surface and internal types. To achieve this we need to create a connection between `LT` and `LineageTransform` type families by showing the compiler that for all types a and k results of reducing `LineageTransform (Rep a) (Rep k)` and `Rep (LT a k)` are equal. This equality means that conversion to internal type representations commutes with the lineage type translation. We need to show this equality for every new surface type a. The crucial step in achieving this is creating a new type class `QLT` and modifying `LT` to become an associated type family of `QLT`:

```
class (QA a) => QLT a where
    type family LT a k
    ltEq :: Proxy a -> Proxy k ->
            LineageTransform (Rep a) (Rep k) :~: Rep (LT a k)
```

where `:~:` represents type equality in Haskell. Instances of `QLT` for primitive types are provided by the library. For user-defined types they are derived automatically using Template Haskell, so the user never sees these instances and needs not to be concerned with them. The `ltEq` function takes two proxy arguments[5] and delivers the required proof of equality between `LineageTransform (Rep a) (Rep k)` and `Rep (LT a k)`. We can then use this proof together with GHC's type casting mechanism in the implementation of the `lineage` library function to perform a safe type cast between lineage-annotated surface types and internal types:

```
lineage :: forall a k. QLT a => Proxy k -> Q a -> Q (LT a k)
lineage pk (Q exp) = let pa = Proxy :: Proxy a in
  Q (castWith (ltEq pa pk) (lineageTransform exp))

lineageTransform :: Exp a -> Exp (LineageTransform a k)
lineageTransform exp = ...
```

The `lineageTransform` function is an internal worker function operating on FL syntax trees and performing the translation defined in Figure 6. It traverses a syntax tree of a query and constructs a new syntax tree of with lineage tracking added.

We should acknowledge that the actual code is a little more complicated than this, e.g. it passes extra arguments to `lineageTransform` so that the k type variable is not ambiguous, but the above discussion outlines the key conceptual obstacle that we needed to overcome to make `lineage` work. Appendix D provides full details.

---

[5] Recall that proxy arguments are needed wherever type variables appear only under type family applications. This is the case with a and k arguments here.

### 4.4.2 Implementing lineage query translation

Implementation of `lineageTransform` also posed a challenge. A problem arises from the `Reify` constraints placed on `VarE` and `ListE` data constructors in `Exp`. When a type class constraint is placed on a data constructor it acts like an extra implicit argument. At every use site of such a constructor, the compiler determines the instantiation of a constraint and replaces it with a *dictionary*, a record that provides implementations of type class methods for a concrete instantiated type. But when we build a syntax tree inside `lineageTransform`, the values stored inside newly created `VarE` and `ListE` expressions do not determine the indexing types and the compiler cannot instantiate required `Reify` dictionaries. Thus we have to help the compiler by guiding type inference. But type class dictionaries are implicit – they do not appear in the source code and cannot be manipulated. Our solution is to modify the definition of `Exp` data type by turning ambiguous `Reify` constraints on `ListE` and `VarE` data constructors into explicit `Type` arguments that we can manipulate:

```
data Exp a where
    BoolE       :: Bool -> Exp Bool
    ListE       :: Type a -> [Exp a] -> Exp [a]
    TableE      :: Reify a => Table -> Exp [a]
    AppE        :: Fun a b -> Exp a -> Exp b
    VarE        :: Type a -> Integer -> Exp a
    TupleConstE :: TupleConst a -> Exp a
```

To modify these explicit `Type` arguments during lineage transformation we need a function that implements the type translation in Figure 7 on `Type` expressions:

```
typeLT :: Type a -> Type k -> Type (LineageTransform a k)
typeLT (BoolT)     _   = BoolT
typeLT (ListT lt) kt = ListT (TupleT (Tuple2T (typeLT lt kt)
                              (ListT (TupleT (Tuple2T TextT kt)))))
typeLT (TupleT (Tuple2T a b)) k = TupleT (Tuple2T (typeLT a k)
                                              (typeLT b k))
-- more equations for primitive types and tuples
```

Here we take an expression of `Type a` (type of things being annotated) and `Type k` (type of keys) and return a term that encodes `Type (LineageTransform a k)`, i.e. a type of a with lineage annotation attached. Using the `LineageTransform` type family in the return type creates a connection between term-level transformation performed on CL types by `typeLT` and type-level transformation on FL types. This is the third time we have to implement lineage type translation — an unsatisfactory state of affairs from a perspective of software engineering.

   We face the final problem when rewriting `TableE` expressions, where each row has to receive a lineage annotation containing the row's key. Projecting a key requires constructing an expression for tuple projection. For example in the `"agencies"` table, where the first of four columns is the key we would need to construct an expression `AppE Tup4_1 row`, where `Tup4_1` is a built-in function for projecting first component of a 4-tuple and `row` is a variable that represents a table row. The types of the involved expressions are:

```
Tup4_1 :: Fun (a,b,c,d) d
AppE   :: Fun a b -> Exp a -> Exp b
```

In order for our projection to typecheck, `row` would need to have type `Exp (a,b,c,d)`, so that it matches the type of argument required by `AppE`. However, `row` has the type `Exp e`, which is a direct consequence of `TupleConstE` data constructor losing information about the index of wrapped `TupleConst` (recall Section 4.2). Thus, such a projection does not typecheck and cannot be implemented in this way. How come we did not face this problem when implementing where-provenance? After all, annotating `TableE` expressions with where-provenance also required projecting the row keys. The difference comes from representation of types in FL and CL. In FL, used during lineage transformation, types are encoded using GADT indices and by relying solely on Haskell's type inference we cannot express well-typedness of a key projection. But a CL syntax tree constructed during where-provenance transformation is an ordinary ADT with types stored explicitly in expressions. This allowed us to implement where-provenance key projection by manually constructing type that describes it, just in the same way we use `typeLT` to explicitly manipulate types during lineage transformation above.

A solution we adopt in this case is to extend existing DSH table definitions to contain a function for projecting the primary key:

```
agenciesL :: Q [Agency]
agenciesL = table "agencies"
                  [ "a_id", "a_name", "a_based_in", "a_phone" ]
                  a_idQ  -- key projection function
                  (TableHints [Key ["a_id"]] NoProvenance)
```

To ensure that the type of keys returned by a projection function matches the type of key specified by the programmer for lineage tracking we perform a runtime type equality test. This is not a usual thing in Haskell, which erases types during compilation thus requiring us to rely on even more extensions beyond the language standard.

Appendix D shows full implementation of `lineageTransform` function for table expressions, with commentary.

### 4.4.3 Lineage implementation conclusions

Implementing lineage was far more challenging than implementing where-provenance. It was to be expected that more code will be needed, since lineage transformation requires rewriting the whole syntax tree. But to implement lineage we also had to employ a number of advanced Haskell techniques to convince the type checker that our implementation is indeed well-typed. The problems arose from the fact that DSH embeds its typing rules inside Haskell's type system by means of GADT indexing. This is often considered to be an advantage of EDSLs since it allows to piggyback on host language's type checker, relieving language designer from having to implement their own. As shown above we can quickly reach the limits of this technique. When this happens we are forced to implement those fragments of our EDSL's type checker that cannot be handled by embedding inside the host language type system.

## 5 Conclusions

In this paper we have taken provenance tracking techniques developed initially as ad hoc database or language extensions and implemented them inside the Database-Supported Haskell library, practically reproducing earlier findings in a different context. We provide a working implementation [15] that, unlike earlier work in Links language [11], fulfils all safety requirements for provenance [3]. Our work demonstrates that provenance tracking does not have to be built into a language or database implementation, but can be provided as a library instead. This is an important step towards supporting provenance tracking for scientific database systems written in mainstream programming languages.

Our work is intended as an exploration of the provenance-as-library idea in a Haskell setting and the implementation acts as a proof of concept. We demonstrated that implementing provenance as part of an EDSL carries its unique challenges related to embedding query typing rules in host language's type system. When our embedding came up against the limitations of Haskell's type inference mechanisms we had to implement our own type checking for some fragments of the lineage transformation. This was possible in Haskell thanks to type families providing expressive power required to convince the type checker that lineage transformation produces well-typed syntax trees. Advanced features of Haskell were necessary in our implementation because DSH encodes typing invariants as GADT indices, and we needed to ensure that these invariants were maintained by our transformations. While our approach shows that these features of Haskell are *sufficient*, it remains an open question what language features are *necessary* to support our approach. We conjecture that a basic requirement for implementing provenance as a library is the ability to inspect and transform query syntax trees in a type-preserving way. In case of DSH this is achieved by providing a custom desugaring of surface Haskell syntax. Further experimentation with other languages and meta-programming techniques (e.g. in F# or Scala) is necessary to determine minimal sets of language features needed to implement provenance as a library approach.

We have not conducted a detailed performance evaluation of our approach, but on tested examples it yields similar queries to those in Links, so should have similar performance to that reported by Fehrenbach and Cheney [11].

Future work on this topic includes provenance tracking for non-monotone operators, which constitute a significant chunk of DSH's built-in primitive functions. We would also like to explore composition of where-provenance and lineage in a single query. There seems to be no theoretical reason that would prevent us from having both forms of provenance in a single query, but the problem remains an engineering challenge.

**Acknowledgements**   We thank Alexander Ulrich and Torsten Grust, the authors of the DSH library, for helping us understand the library implementation. This work was supported by ERC Consolidator Grant Skye (grant number 682315).

## A   Basics of Haskell

Below we give a quick tour of some of Haskell's standard features, defined in the Haskell language report [17]), that set it apart from the majority of mainstream programming languages.

### A.1  Standard Prelude

The Haskell Language Report [17] defines what is called a *standard Prelude* - a module that provides definitions of basic data types and functions that constitute an essential core of Haskell language. The Standard Prelude is implicitly imported into every Haskell module (source file).

### A.2  Algebraic Data Types

At the centre of Haskell programming, and many other functional languages, are Algebraic Data Types (or ADTs for short). Here is a simple example that should be immediately obvious to any programmer:

```
data Bool where
  False :: Bool
  True  :: Bool
```

This definition introduces type `Bool` representing logical values. Expressions of type `Bool` take one of two values: either `False` or `True`. In Haskell lingo `Bool` is called a *type constructor*, while `False` and `True` are known as *data constructors*.

Here is a more complicated example that defines a type of lists:

```
data List a where
  Nil  :: List a
  Cons :: a -> List a -> List a
```

This data type is *parametrized* by a *type variable* a that is instantiated to a concrete type, e.g. `Bool`, when values are created. `Nil` data constructor creates an empty list. `Cons` appends a single element to the front of already existing list (note that this makes `List` a recursive data type). So a list containing booleans `True` and `False` would be written as `Cons True (Cons False Nil)`. In fact, lists are so common in Haskell that they have a special notation. `Nil` is written as `[]` and `Cons` is written as an infix operator `:`. So the list above could be written as `True : False : []` or, even more succinctly, `[True, False]`.

Haskell also allows to define ADTs using so-called *record syntax*:

```
data Employee = Employee
    { name   :: Text
    , salary :: Integer }
```

The above declaration creates a data type `Employee` that has a single constructor (also named `Employee`) with two fields stored in it (just like `Cons` stored two values. These values can be accessed using `name` and `salary` accessor functions, just like in any other language with records.

### A.3  Type Classes

Another distinguishing feature of Haskell is its *type class* facility, a mechanism for ad-hoc polymorphism. Type classes allow to declare a set of abstract operations, with implementations for concrete data types provided by *type class instances*. Type classes are open, meaning that new instances can be freely added by the programmer. A commonly encountered type class in Haskell is `Eq`:

```
class Eq a where
    (==) :: a -> a -> Bool
    (/=) :: a -> a -> Bool
    x /= y = not (x == y)
```

It defines equality and inequality between values of some abstract type a. The inequality operator `/=` is given a *default implementation* defined as a negation of equality. This way, a programmer writing an instance of `Eq` needs only to provide a definition of `==` operator (though she may override the default implementation of `/=` if she wishes):

```
instance Eq Bool where
    True  == True  = True
    False == False = True
    _     == _     = False
```

Having defined a type class, a programmer can request that a type parameter of a polymorphic function is an instance of a given type class:

```
elem :: Eq a => a -> [a] -> Bool
```

This type signature for `elem` function says that in order to test whether a value of type a is an element of a given list of as, a needs to be an instance of `Eq` type class so that it can be tested for equality. The `(Eq a)` part of the signature before `=>` arrow is called a type class constraint.

### A.4  List comprehensions

Lists come with another convenient notation that allows to construct new lists from existing ones. This notation is just a syntactic sugar for more primitive functions and operators defined in the standard Prelude. It is modelled on *set comprehension* notation from mathematics. It is best explained by example:

```
[ x * x | x <- [1..10], even x]
```

List comprehension above takes elements from a list of 1 to 10, filters out the odd numbers, and then squares each number that's left, producing list `[4,16,36,64,100]` as a result. In this notation `x <- [1..10]` is called a *generator* and `even x` is called a *guard*.

List comprehensions are just syntactic sugar for functions defined in the standard Prelude. Most importatly, iteration over a generator desugars into a call to `concatMap`, so `[... | x <- xs, ...]` desugars into `concatMap (\x -> ...) xs`.

## B  Examples of DSH queries with provenance tracking

This appendix contains complete examples of DSH queries with where-provenance and lineage tracking. Examples include data definitions, calls to Template Haskell functions generating boilerplate, table definitions and the actual query. Note that code in this section is prettified in few places to elide unimportant technical details.

### B.1  Where-provenance tracking

Listing 1 presents an example of where-provenance tracking. The code begins with declaring `AgencyWP` (lines 1-5) and `ExternalTour` (lines 10-15) data types to represent rows of database tables. In the declaration of `Agency` the `a_phone` field is marked with where-provenance tracking using `Integer` keys. We generate DSH boilerplate by calling functions `deriveDSH` and `generateTableSelectors` (lines 7-8 and 17-18), both written using Template Haskell. We follow with table definitions (lines 20-30). For the `"agencies"` table we declare where-provenance tracking for the `a_phone` column. The declaration of the `externaltours` table specifies the `NoProvenance` hint, since we do not track where-provenance for any of its columns. Notice that both table declarations require specifying functions for projecting primary keys (`a_idQ` on line 23, `et_idQ` on line 29), even though these functions are only used during lineage transformation.

### B.2  Lineage tracking

Listing 2 presents an example of lineage tracking. The code is very similar to the previous example with where-provenance, but there are some key differences. The declaration of `Agency` data type obviously does not contain a where-provenance annotation on `a_phone` field (line 5). Correspondingly, there is no provenance hint in the `agencies` table declaration (line 24). This means that if a programmer wants to add lineage tracking to their queries they do not need to modify any of the existing declarations. Unlike with where-provenance, where we had to modify the query to account for provenance, here we have two queries. One is the original query `q1` (lines 32-35). We also have a second query `q1''` with lineage tracking. It calls the library function `lineage` and passes it a proxy argument to specify the type of keys and the `q1` query to be extended with lineage tracking. The type signature of `q1''` (line 37) uses the `LT` type family, but we could also write the signature as

■ **Listing 1** Where-provenance tracking example

```
1  data AgencyWP = AgencyWP
2    { a_id       :: Integer
3    , a_name     :: Text
4    , a_based_in :: Text
5    , a_phone    :: WhereProv Text Integer }
6
7  deriveDSH            ''AgencyWP
8  generateTableSelectors ''AgencyWP
9
10 data ExternalTour = ExternalTour
11   { et_id        :: Integer
12   , et_name      :: Text
13   , et_destination :: Text
14   , et_type      :: Text
15   , et_price     :: Integer }
16
17 deriveDSH            ''ExternalTour
18 generateTableSelectors ''ExternalTour
19
20 agenciesWP :: Q [AgencyWP]
21 agenciesWP = table "agencies"
22               [ "a_id", "a_name", "a_based_in", "a_phone" ]
23               a_idQ
24               (TableHints [Key ["a_id"]] (WhereProvenance ["a_phone"]))
25
26 externaltours :: Q [ExternalTour]
27 externaltours = table "externaltours"
28     [ "et_id", "et_name" , "et_destination" , "et_type" , "et_price" ])
29     et_idQ
30     (TableHints [Key ["et_id"]] NoProvenance)
31
32 q1 :: Q [(Text, WhereProv Text Integer)]
33 q1 = [ tup2 (et_nameQ et) (a_phoneQ a)
34      | a  <- agenciesWP, et <- externaltours
35      , a_nameQ a  == et_nameQ et, et_typeQ et == "boat" ]
```

**■ Listing 2** Lineage tracking example

```
1  data Agency = Agency
2    { a_id       :: Integer
3    , a_name     :: Text
4    , a_based_in :: Text
5    , a_phone    :: Text }
6
7  deriveDSH            ''Agency
8  generateTableSelectors ''Agency
9
10 data ExternalTour = ExternalTour
11   { et_id        :: Integer
12   , et_name      :: Text
13   , et_destination :: Text
14   , et_type      :: Text
15   , et_price     :: Integer }
16
17 deriveDSH            ''ExternalTour
18 generateTableSelectors ''ExternalTour
19
20 agencies :: Q [Agency]
21 agencies = table "agencies"
22             [ "a_id", "a_name", "a_based_in", "a_phone" ]
23             a_idQ
24             (TableHints [Key ["a_id"]] NoProvenance)
25
26 externaltours :: Q [ExternalTour]
27 externaltours = table "externaltours"
28      [ "et_id", "et_name" , "et_destination" , "et_type" , "et_price" ])
29      et_idQ
30      (TableHints [Key ["et_id"]] NoProvenance)
31
32 q1 :: Q [(Text, Text)]
33 q1 = [ tup2 (et_nameQ et) (a_phoneQ a)
34      | a  <- agencies, et <- externaltours
35      , a_nameQ a  == et_nameQ et, et_typeQ et == "boat" ]
36
37 q1'' :: Q (LT [(Text, Text)] Integer)
38 q1'' = lineage (Proxy :: Proxy Integer) q1
```

[Lineage (Text, Text) Integer]. This is the type the call to `LT` reduces to and it does not matter whether we use a type family or write the correct type by ourselves.

## C   Examples of lineage transformation

In this appendix we demonstrate how lineage transformation rules from Figure 6 work in practice.

### C.1   Example 1

We begin with a simple example:

```
q0 :: Q [Text]
q0 = [ a_nameQ a | a <- agencies ]
```

This code projects names of all the travel agencies. Its desugaring into our core language looks like this:

**concatMap** $(\lambda a. [\texttt{a\_nameQ}\ a])\ \texttt{agencies}$

where `constant-width font` denotes identifiers coming from the source code. We begin by applying equation for transforming **concatMap** to arrive at an intermediate result:

**concatMap** $(\lambda x.\ \textbf{map}\ (\lambda z.\ z.\textbf{data}^{z.\textbf{prov} \oplus x.\textbf{prov}})\ (\mathfrak{L}_\theta((\lambda a. [\texttt{a\_nameQ}\ a])(x.\textbf{data}))))$

$(\mathfrak{L}_\theta(\texttt{agencies}))$

There are now two recursive calls to lineage transformation $\mathfrak{L}_\theta$. In the first call we begin by simplifying the argument by $\beta$-reduction and then apply equation for transforming list literals. In the second call to $\mathfrak{L}_\theta$ we apply the equation for transforming table declarations, annotating every row in the table with lineage transformation. We arrive at the final transformed query:

**concatMap** $(\lambda x.\ \textbf{map}\ (\lambda z.\ z.\textbf{data}^{z.\textbf{prov} \oplus x.\textbf{prov}})\ (\textbf{map}\ (\lambda x.x^{\perp})\ [\texttt{a\_nameQ}\ (x.\textbf{data})]))$

$(\textbf{map}\ (\lambda x.x^{(\text{``agencies''}, \phi(x))})\ \texttt{agencies})$

### C.2   Example 2

We now turn to performing lineage transformation on an example query used throughout the paper:

```
q1 :: Q [(Text, Text)]
q1 = [ tup2 (et_nameQ et) (a_phoneQ a)
     | a  <- agencies
     , et <- externaltours
     , a_nameQ a == et_nameQ et
     , et_typeQ et == "boat" ]
```

We begin by desugaring this query into our core calculus:

**concatMap** ($\lambda a$.
  **concatMap** ($\lambda et$.
   **concatMap** ($\lambda ()$.
    **concatMap** ($\lambda ()$. ($\texttt{et\_nameQ}\, et, \texttt{a\_phoneQ}\, a$))
    (**guard** ($\texttt{et\_typeQ}\, et\, \texttt{==}\, \texttt{"boat"}$)))
   (**guard** ($\texttt{a\_nameQ}\, a\, \texttt{==}\, \texttt{et\_nameQ}\, et$)))
  $\texttt{externalTours}$)
 $\texttt{agencies}$

One important thing about the above desugaring is that guards in the original list comprehensions got translated into a **concatMap** over a **guard** expressions. Recall that a **guard** built-in function takes a boolean expression as an argument and returns a singleton list containing a unit () if the boolean argument reduces to true or an empty list if it reduces to false. Lambda functions passed to each of two **concatMap**s that iterate over the list returned by **guard** don't bind any variables, but rather pattern match on a unit – the only possible value here. Notice also that the surface function $\texttt{tup2}$ operating on arguments wrapped in $\texttt{Q}$ gets translated to a standard tuple constructor in our core calculus.

We now proceed by applying **concatMap** transformation equation to the outermost **concatMap** expression. We immediately simplify call to $\mathfrak{L}_\theta(\texttt{agencies})$, because we already know what the result is from the previous example:

**concatMap** ($\lambda x$. **map** ($\lambda z.\, z.\textbf{data}^{z.\textbf{prov} \oplus x.\textbf{prov}}$)
  $\mathfrak{L}_\theta$(**concatMap** ($\lambda et$.
   **concatMap** ($\lambda ()$.
    **concatMap** ($\lambda ()$. ($\texttt{et\_nameQ}\, et, \texttt{a\_phoneQ}\, x.\textbf{data}$))
    (**guard** ($\texttt{et\_typeQ}\, et\, \texttt{==}\, \texttt{"boat"}$)))
   (**guard** ($\texttt{a\_nameQ}\, x.\textbf{data}\, \texttt{==}\, \texttt{et\_nameQ}\, et$)))
  $\texttt{externalTours}$)
($\textbf{map}$ ($\lambda x.x^{(\text{``}agencies\text{''}, \phi(x))}$) $\texttt{agencies}$)

We repeat the same transformation for **concatMap** over $\texttt{externalTours}$ table:

**concatMap** $(\lambda x.\ \mathbf{map}\ (\lambda z.\ z.\mathbf{data}^{z.\mathbf{prov}\oplus x.\mathbf{prov}})$
  $(\mathbf{concatMap}\ (\lambda y.\ \mathbf{map}\ (\lambda z.\ z.\mathbf{data}^{z.\mathbf{prov}\oplus y.\mathbf{prov}})$
   $\mathfrak{L}_\theta(\mathbf{concatMap}\ (\lambda().$
    $\mathbf{concatMap}\ (\lambda().\ (\texttt{et\_nameQ}\ y.\mathbf{data}, \texttt{a\_phoneQ}\ x.\mathbf{data}))$
    $(\mathbf{guard}\ (\texttt{et\_typeQ}\ y.\mathbf{data}\ ==\ \texttt{"boat"})))$
    $(\mathbf{guard}\ (\texttt{a\_nameQ}\ x.\mathbf{data}\ ==\ \texttt{et\_nameQ}\ y.\mathbf{data}))))$
  $(\mathbf{map}\ (\lambda x.x^{(\text{``externalTours''}, \phi(x))})\ \texttt{externalTours}))$
$(\mathbf{map}\ (\lambda x.x^{(\text{``agencies''}, \phi(x))})\ \texttt{agencies})$

We now transform **concatMap** over the first **guard** expression, applying the appropriate equation for **guard** and immediately simplifying second call to $\mathfrak{L}_\theta$:

**concatMap** $(\lambda x.\ \mathbf{map}\ (\lambda z.\ z.\mathbf{data}^{z.\mathbf{prov}\oplus x.\mathbf{prov}})$
  $(\mathbf{concatMap}\ (\lambda y.\ \mathbf{map}\ (\lambda z.\ z.\mathbf{data}^{z.\mathbf{prov}\oplus y.\mathbf{prov}})$
   $(\mathbf{concatMap}\ (\lambda v.\ \mathbf{map}\ (\lambda z.\ z.\mathbf{data}^{z.\mathbf{prov}\oplus v.\mathbf{prov}})$
    $\mathfrak{L}_\theta(\mathbf{concatMap}\ (\lambda().\ (\texttt{et\_nameQ}\ y.\mathbf{data}, \texttt{a\_phoneQ}\ x.\mathbf{data}))$
    $(\mathbf{guard}\ (\texttt{et\_typeQ}\ y.\mathbf{data}\ ==\ \texttt{"boat"})))$
    $(\mathbf{map}\ (\lambda x.x^\perp)\ (\mathbf{guard}\ (\texttt{a\_nameQ}\ x.\mathbf{data}\ ==\ \texttt{et\_nameQ}\ y.\mathbf{data}))))$
  $(\mathbf{map}\ (\lambda x.x^{(\text{``externalTours''}, \phi(x))})\ \texttt{externalTours})))$
$(\mathbf{map}\ (\lambda x.x^{(\text{``agencies''}, \phi(x))})\ \texttt{agencies})$

We repeat the same step for the second **concatMap** over **guard**:

**concatMap** $(\lambda x.\ \mathbf{map}\ (\lambda z.\ z.\mathbf{data}^{z.\mathbf{prov}\oplus x.\mathbf{prov}})$
  $(\mathbf{concatMap}\ (\lambda y.\ \mathbf{map}\ (\lambda z.\ z.\mathbf{data}^{z.\mathbf{prov}\oplus y.\mathbf{prov}})$
  $(\mathbf{concatMap}\ (\lambda v.\ \mathbf{map}\ (\lambda z.\ z.\mathbf{data}^{z.\mathbf{prov}\oplus v.\mathbf{prov}})$
   $(\mathbf{concatMap}\ (\lambda w.\ \mathbf{map}\ (\lambda z.\ z.\mathbf{data}^{z.\mathbf{prov}\oplus w.\mathbf{prov}})$
    $\mathfrak{L}_\theta(\texttt{et\_nameQ}\ y.\mathbf{data}, \texttt{a\_phoneQ}\ x.\mathbf{data}))$
   $(\mathbf{map}\ (\lambda x.x^\perp)\ (\mathbf{guard}\ (\texttt{et\_typeQ}\ y.\mathbf{data}\ ==\ \texttt{"boat"})))))$
  $(\mathbf{map}\ (\lambda x.x^\perp)\ (\mathbf{guard}\ (\texttt{a\_nameQ}\ x.\mathbf{data}\ ==\ \texttt{et\_nameQ}\ y.\mathbf{data})))))$
  $(\mathbf{map}\ (\lambda x.x^{(\text{``externalTours''}, \phi(x))})\ \texttt{externalTours})))$
$(\mathbf{map}\ (\lambda x.x^{(\text{``agencies''}, \phi(x))})\ \texttt{agencies})$

We are left with performing lineage transform for the innermost tuple expression. We inject $\mathfrak{L}_\theta$ under tuple constructor and apply it to every component. Applications of `et_nameQ` and `a_phoneQ` are treated as constants and remain unchanged. We thus arrive at the final transformed expression:

■ **Listing 3**  Implementation of `lineage` function

```
1  lineage :: forall a k. ( QLT a, QA k, Typeable (Rep k) )
2          => Proxy k -> Q a -> Q (LT a k)
3  lineage pk (Q e) = let pa = Proxy :: Proxy a in
4      Q (castWith (apply Refl (ltEq pa pk))
5          (runLineage (lineageTransform (reifyTy :: Type (Rep a))
6                                        (reifyTy :: Type (Rep k)) e)))
```

$$\textbf{concatMap}\ (\lambda x.\ \textbf{map}\ (\lambda z.\ z.\textbf{data}^{z.\textbf{prov} \oplus x.\textbf{prov}})$$
$$(\textbf{concatMap}\ (\lambda y.\ \textbf{map}\ (\lambda z.\ z.\textbf{data}^{z.\textbf{prov} \oplus y.\textbf{prov}})$$
$$(\textbf{concatMap}\ (\lambda v.\ \textbf{map}\ (\lambda z.\ z.\textbf{data}^{z.\textbf{prov} \oplus v.\textbf{prov}})$$
$$(\textbf{concatMap}\ (\lambda w.\ \textbf{map}\ (\lambda z.\ z.\textbf{data}^{z.\textbf{prov} \oplus w.\textbf{prov}})$$
$$(\texttt{et\_nameQ}\ y.\textbf{data}, \texttt{a\_phoneQ}\ x.\textbf{data}))$$
$$(\textbf{map}\ (\lambda x.x^{\perp}\ (\textbf{guard}\ (\texttt{et\_typeQ}\ y.\textbf{data} == \texttt{"boat"})))))$$
$$(\textbf{map}\ (\lambda x.x^{\perp}\ (\textbf{guard}\ (\texttt{a\_nameQ}\ x.\textbf{data} == \texttt{et\_nameQ}\ y.\textbf{data})))))$$
$$(\textbf{map}\ (\lambda x.x^{(\text{``externalTours''}, \phi(x))}\ \texttt{externalTours})))$$
$$(\textbf{map}\ (\lambda x.x^{(\text{``agencies''}, \phi(x))}\ \texttt{agencies})$$

This expression, when compiled further to SQL and executed on the database, will yield a result shown in Figure 3.

## D  Lineage implementation details

In this appendix we fill in the missing technical details of `lineage` and `lineageTransform` implementation that we elided in Sections 4.4.1 and 4.4.2. This code uses monads and the **do**-notation, which were not covered in Appendix A, but are still one of the standard features of Haskell.

Listing 3 shows the the implementation of `lineage` function, exposed to the user as the library interface. The type signature (lines 1-2) contains two type variables: a for the type being annotated with lineage and k for the type of table keys. Line 1 also specifies type class constraints: `QLT a` says that for type a there must be a defined way to convert between surface and internal lineage-annotated types; `QA k` says that the type of keys must be representable internally in DSH; `Typeable (Rep k)` says that it must be possible to test runtime type equality for internal representation of keys – we will need it in the implementation of `lineageTransform` below. The `lineage` function takes as its argument a query of type `Q a` and returns a transformed query of type `Q (LT a k)`, containing lineage tracking. Since type variable k appears only under type family applications (`Rep k` on line 1, `LT a k` on line 2) and type class constraints (`QA k` on line 1) we need to pass an extra proxy argument of type `Proxy k`.

■ **Listing 4**  Implementation of `reifyTy` function

```
1  reifyTy :: forall a. Reify a => Type a
2  reifyTy = reify (undefined :: a)
```

■ **Listing 5**  Fragment of `lineageTransform` implementation handling `TableE` case.

```
1  lineageTransform :: Typeable k
2                   => Type a -> Type k -> Exp a
3                   -> Compile (Exp (LineageTransform a k))
4  lineageTransform tyA tyK tbl@(TableE (TableDB name _ _) keyProj) = do
5    let keyEquality :: (Typeable k1, Typeable k2)
6                    => Type k1 -> (Integer -> Exp k2) -> Maybe (k1 :~: k2)
7        keyEquality _ _ = eqT
8    case keyEquality tyK keyProj of
9      Just Refl -> do
10       let lam t a = lineageE (VarE t a)
11                       (lineageAnnotE tyK (pack name) (keyProj a))
12       return (AppE Map (TupleConstE (Tuple2E
13                   (LamE (elemTy tyA) (lam (typeLT reifyTy tyK))) tbl)))
14     Nothing ->
15       error "Type of table key does not match type of lineage key"
```

The most important part of `lineage` body (lines 3-6) is the call to `lineageTransform` worker. We pass three arguments to `lineageTransform`. The first two are explicit type representations of internal types of `a` and `k`. These arguments are required for explicit type manipulations described in Section 4.4.2 and we construct them using the `reifyTy` helper function (Listing 4), which can create an explicit type representation `Type a` for any type `a` that is an instance of `Reify` type class. The third argument to `lineageTransform` is the query syntax tree of type `Exp (Rep a)`. The call to `lineageTransform` is wrapped in `runLineage` function, which runs computations inside the `Compile` monad and returns their result. `Compile` is a simple state monad used to generate fresh variable names during query compilation.

Let's think for a moment about the types at this point. We need to build a result of type `Q (LT a k)`. By definition of `Q` newtype on page 17 we need an expression of type `Exp (Rep (LT a k))`. But what we actually get from the call to `lineageTransform` is of a different type: `Exp (LineageTransform (Rep a) (Rep k))`. We expect these types to be the same, but the compiler cannot figure this out on its own. That's where our `QLT` type class becomes essential, as it allows us to use GHC's safe type casting mechanism. By calling `ltEq` with appropriate proxy arguments (line 4) we obtain equality between `Rep (LT a k)` and `LineageTransform (Rep a) (Rep k)`. We now need to inject this equality under `Exp` type constructor. This is done with the call to `apply`, where the `Refl` argument is a proof that type constructor `Exp` is equal to itself. We then perform the cast using `castWith` function and wrap the result in `Q` (line 4).

Listing 5 shows the implementation of lineage transformation for `TableE` expressions. Lines 1-3 contain the type signature. As already explained, `lineageTransform` takes as arguments explicit `Type` representations of types `a` and `k` as well as an `Exp a` tree to transform (line 2). (Note: a and k types in `lineageTransform` signature correspond to `Rep a` and `Rep k` in the `lineage` function.) The result of `lineageTransform` is described in terms of `LineageTranform` type family and returned inside `Compile` monad (line 3).

On line 4 we pattern match on `TableE` constructor and bind the names of the arguments. The `tyA` and `tyK` are internal representation of Haskell types a and k, respectively.

To transform a table expression we need to make sure that the type of table key matches the type of key specified for tracking lineage. Since in DSH queries are compiled at runtime, we have to test for type equality at runtime. We create a helper function `keyEquality` (lines 5-7) that takes arguments parametrized by two distinct type variables k1 and k2. Both of these types have to belong to a `Typeable` type class, which enables runtime equality testing. Our function returns a `Maybe`: either a proof of type equality wrapped in a `Just` (if types k1 and k2 are definitionally equal) or `Nothing` (if types k1 and k2 are distinct). If the latter happens `lineageTransform` fails with an error (lines 14-15). If the types are equal we construct a syntax tree according to equation in Figure 6 (lines 12-13). We rely on several helper definitions. `lam` (lines 10-11) constructs a lambda that appends lineage to a variable. The `lineageE` and `lineageAnnotE` construct `Exp` expressions corresponding to internal representations of `Lineage` and `LineageAnnot` data types. On line 11 we project the primary key of a row, passing the lambda variable as an argument. The `elemTy` helper (line 13) takes a list type and projects out the type of list elements. On line 13 we also call the `typeLT` function responsible for implementing lineage type translation defined in Figure 7. This is the explicit manipulation of type representations described in Section 4.4.2.

## About the authors

**Jan Stolarek** is a Research Associate in the Laboratory for Foundations of Computer Science, University of Edinburgh. Email: jan.stolarek@ed.ac.uk

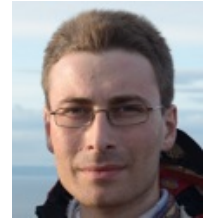

**James Cheney** is a Reader in the Laboratory for Foundations of Computer Science, University of Edinburgh. Email: jcheney@inf.ed.ac.uk

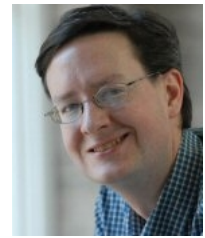